\documentclass[preprint,prd,nofootinbib,tightenlines,superscriptaddress]{article}

\usepackage{color}
\usepackage{graphicx}
\definecolor{red}{rgb}{1,0,0}
\def\lesssim{\ \hbox{\raise 2pt \hbox{$<$} \kern -13pt
                     \lower 3pt \hbox{$\sim$}}\ }
\def\greatersim{\ \hbox{\raise 2pt \hbox{$>$} \kern -13pt
                     \lower 3pt \hbox{$\sim$}}\ }

\input epsf.tex
\def\desepsf(#1 width #2){\epsfxsize=#2 \epsfbox{#1}}

\usepackage{t1enc}
\usepackage{setspace}
\usepackage[english]{babel}
\usepackage[latin2]{inputenc}
\usepackage{epsfig}
\usepackage{fancyhdr}
\usepackage[abs]{overpic}
\usepackage[normalsize,bf]{caption2}
\usepackage{afterpage}
\usepackage{fancybox}
\usepackage{slashed}
\usepackage{./mcite}
\usepackage{dsfont}

\usepackage{amsmath,bm}

\begin{document}
\title{Estimation of saturation and coherence effects in the KGBJS equation -- a non-linear CCFM equation}
\author{M. De\'ak$$\\
\quad\\
{\it Departamento de F\'isica de Part\'iculas, Facultade de F\'isica,}\\ {\it Universidade de Santiago de Compostela, Campus Sur,}\\{\it 15706 Santiago de Compostela, Spain}}

\maketitle

\abstract

We solve the modified non-linear extension of the CCFM equation -- KGBJS equation~\mcite{Kutak:2012yr,*Kutak:2011fu} -- numerically for certain initial conditions and compare the resulting gluon Green functions with those obtained from solving the original CCFM equation and the BFKL and BK equations for the same initial conditions. We improve the low transversal momentum behaviour of the KGBJS equation by a small modification.

\section{Introduction}

The BFKL equation~\mcite{BFKL1,*BFKL2,*BFKL3,*BFKL4} seems to be the appropriate effective theory for describing high energy initial state radiation in the kinematic region where the Mandelstam invariants $t$ of the momentum exchange and $s$ of the total scattering momentum are strongly ordered: $\Lambda_{QCD}^2\ll |t|\ll s$. It is well known that the relevant phase space region is very important at the LHC, but signs of its importance were already observed at the HERA~\mcite{Aaron:2011ef,Hentschinski:2012kr} and other colliders. 

However, the BFKL equation predicts too strong rise of the cross section for decreasing ratio of $|t|/s$ violating the Froisart bound and thus the unitarity. To solve the problem of the rise of the BFKL cross section different kinds of corrections, from careful consideration of the kinematical constraint~\mcite{Beuf:2011st,*Beuf:2011he}, inclusion of NLO terms~\mcite{Fadin:1998py,*Bartels:2007ms} to inclusion of non-linear terms, were suggested and studied. A non-linear extensions of the equation was proposed~\mcite{Mueller:1990er} to take into account merging of over-populated gluons and thus damping the growth of the gluon density and consequently the cross section. The BK equation~\mcite{Kovchegov:1999yj,*Kovchegov:1999ua,*Balitsky:1995ub} is one of such extensions of the BFKL equation. The growth of the solution of the BK equation is suppressed compared to the solution of the linear equation. Saturation of the gluon density function is observed~\mcite{GolecBiernat:2001if,Marquet:2005zf,Kutak:2011rb,Avsar:2011ds}.

The BFKL and also the BK equations are only suitable to describe the inclusive cross section. Inclusion of coherence effects was proposed to extend the validity of latter equations for exclusive final states. An equation which includes the BFKL kernel plus the coherence effects and at the same time it interpolates between the BFKL and the DGLAP~\mcite{DGLAP1,*DGLAP2,*DGLAP3,*DGLAP4} approximations is the CCFM equation~\mcite{CCFM1,*CCFM2,*CCFM3,*CCFM4}.

An interesting question raises: How does a corresponding non-linear equation, a non-linear extension of the CCFM equation, look like and how does the coherence requirement interplay with the saturation constraint?

As we mentioned before, the CCFM equation allows for study of exclusive final states.  Small-$x$ effects in exclusive observables has been studied using Monte Carlo programs such as {\sc Cascade}~\mcite{CASCADE1,*CASCADE2,Jung:2010si}. By including saturation effects into the CCFM equation, it would therefore possible to study effects of saturation on exclusive states like jet final states, jet multiplicities and energy flows. Attempts were done to include saturation effects by a simple cut-off~\mcite{Avsar:2009pf,*Avsar:2010ia} and also by a sophisticated dipole model~\mcite{Flensburg:2011kk}. In~\mcite{Kutak:2012qk} a non-linear equation for a factorisable gluon density was derived with a proposal of inclusion of coherence effects. A non-linear extension of the CCFM equation in a simple from was suggested in~\mcite{Kutak:2012yr,*Kutak:2011fu}. The latter proposed non-linear equation was not yet studied in the literature. We will start filling the emerged gap in this publication. After that we will take a look on some of the above mentioned well known equations. Then we will examine closely the new non-linear equation and find non-physical behaviour near the soft cut-off. We will suggest an improvement of the equation and present numerical results.

\section{Equations}\label{sec:equations}

In this section we will shortly introduce equations of our interest.

\subsection{The BFKL and the BK equations in momentum space}\label{sec:BFKL}

\begin{figure}[htb]
\begin{center}
\epsfig{figure=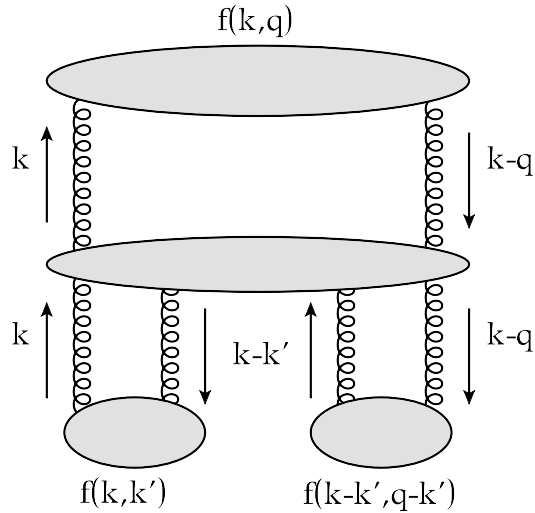,width=7cm,clip}
\caption{Diagramatic representation of the BK equation in momentum space.}\label{fig:BKeq}
\end{center}
\end{figure}

The BK equation in the momentum space, written for the gluon Green function, includes the BFKL equation and a non-linear term with a minus sign in a form of convolution of two gluon Green functions $f(Y,{\bf k},{\bf q})$. The gluon Green function depends on the rapidity $Y$ and two transversal momenta ${\bf k}$ and ${\bf q}$. The non-linear term can be diagramaticaly represented as a merging of the BFKL ladder diagrams (figure~\ref{fig:BKeq}). The BK equation reads~\mcite{Marquet:2005zf,Kutak:2011rb,Avsar:2011ds}
\begin{equation}\label{eq:BKeqnonf}
\begin{split}
\frac{\partial f(Y,{\bf k},{\bf q})}{\partial Y}=\frac{\bar\alpha_S}{\pi}\int\frac{d^2{\bf k}^{\prime}}{({\bf k}-{\bf k}^{\prime})^2}\Bigg\{f(Y,{\bf k}^\prime,{\bf q})&\\-\frac{1}{2}\bigg[\frac{{\bf k}^2}{{{\bf k}^\prime}^2+({\bf k}-{\bf k}^\prime)^2}+&\frac{({\bf q}-{\bf k})^2}{{({\bf q}-{\bf k}^\prime})^2+({\bf k}-{\bf k}^\prime)^2}\bigg]f(Y,{\bf k},{\bf q})\Bigg\}\\
-&\frac{\bar\alpha_S}{2\pi}\int d^2{\bf k}^{\prime}f(Y,{\bf k},{\bf k}^{\prime})f(Y,{\bf k}-{\bf k}^\prime,{\bf q}-{\bf k}^\prime)\;
\end{split}
\end{equation}

(${\bar \alpha}_S=N_C\alpha_S/\pi$).
In the forward limit ${\bf q}\rightarrow 0$ and no impact parameter dependence approximations, the BK equation takes the form:
\begin{equation}\label{eq:BKeq}
\begin{split}
\frac{\partial f(Y,{\bf k})}{\partial Y}=\frac{\bar\alpha_S}{\pi}\int\frac{d^2{\bf k}^{\prime}}{({\bf k}-{\bf k}^{\prime})^2}\Bigg\{f(Y,{\bf k}^\prime)&-\frac{{\bf k}^2}{{{\bf k}^\prime}^2+({\bf k}-{\bf k}^\prime)^2}f(Y,{\bf k})\Bigg\}\\
&-\frac{\bar\alpha_S}{2\pi}\int d^2{\bf k}^{\prime}\delta^{(2)}({\bf k}^\prime)f(Y,{\bf k})f(Y,{\bf k}-{\bf k}^\prime)\;.
\end{split}
\end{equation}

If the Green function grows beyond certain limit, the non-linear term causes a suppression of the growth, because of the negative sign and because of being effectively quadratic in the Green function. This suppression of the cross section was successfully shown in many publications~\mcite{Marquet:2005zf}.

The authors of~\mcite{Kutak:2012yr,*Kutak:2011fu} have shown how to resum the virtual corrections, included in the second term on the right hand side of the equation~\eqref{eq:BKeq}, and obtain this form of the BK equation
\begin{equation}\label{eq:BKresummed}
\begin{split}
f(x,{\bf k}^2)=\tilde f_0({\bf k}^2)+\frac{\bar\alpha_S}{\pi}\int\limits_x^1\frac{dz}{z}\Delta_R(z,{\bf k}^2,\mu)\bigg[\int\frac{d^2{\bf q}}{\pi{\bf q}^2}\theta({\bf q}^2-\mu^2)f(x/z,|{\bf k}+{\bf q}|^2)&\\-f^2(x/z,{\bf k}^2)&\bigg]\;,
\end{split}
\end{equation}

where this time $x$ is the momentum fraction of the proton carried by the parton, ${\bf k}$ is the partons transversal momentum. The function $\tilde f_0({\bf k}^2)$ is the initial condition of the equation. The integration ${\bf q}$ can be interpreted as the transversal momentum of the emitted parton and $z$ can be interpreted as the momentum fraction of the mother parton carried by the recoiling parton.
The Regge form factor $\Delta_R(z,{\bf k}^2,\mu)=\exp\big(-\bar\alpha_S\ln(1/z)\ln({\bf k}^2/\mu^2)\big)$ includes now the resummed virtual corrections. A soft cut-off $\mu$ is introduced.

\subsection{The CCFM equation}

The CCFM equation reads
\begin{equation}\label{eq:IS-CCFM}
\begin{split}
\mathcal{F}(x,{\bf k},p)&=\mathcal{F}_0({\bf k})+{\bar\alpha}_S\int\frac{d^2\bar{\bf q}^\prime}{\bar{\bf q}^{\prime 2}}
\int\limits_{x}^{1-\frac{Q_0}{|{\bar{\bf q}}^{\prime}|}}\frac{dz}{z}\mathcal{F}(x/z,{\bf k}^{\prime},|\bar{\bf q}^{\prime}|)\\&\times\;\theta(p-z|{\bar{\bf q}}^\prime|){\mathcal P}(z,{\bf k},{\bf q})\Delta_S(p,z|\bar{\bf q}^{\prime}|,Q_0)\;,
\end{split}
\end{equation}

with similar meaning of variables $x$, ${\bf k}$, ${\bf q}$ and $z$ as in the previous subsection~\ref{sec:BFKL} and $\mathcal{F}_0({\bf k})$ being now the initial condition. In addition the variable $\bar{\bf q}={\bf q}/(1-z)=({\bf k}^\prime-{\bf k})/(1-z)$ and a new scale $p$ characterising the hard scale or the maximum emission angle in the evolution are introduced.
The Sudakov form factor $\Delta_S(p,(z\bar{\bf q})^2)$ reads
\begin{equation}\label{eq:sud}
\Delta_S(p,z|\bar{\bf q}|,Q_0)=\exp\Bigg(-\int\limits_{(z\bar{\bf q})^2}^{p^2}\frac{d^2{\bf q}^{\prime}}{\pi{{\bf q}^{\prime}}^2}\int\limits_{0}^{1-\frac{Q_0}{|{\bf q}^\prime|}}dz^\prime\frac{\bar\alpha_S}{1-z^\prime}\Bigg)
\end{equation}

and the Non-Sudakov form factor $\Delta_{NS}({\bf k}^2,(z\bar{\bf q})^2)$ is defined as
\begin{equation}\label{eq:nsud}
\Delta_{NS}(z,{\bf k}^2,|{\bf q}|)=\exp\Bigg(-\int\limits_{(z\bar{\bf q})^2}^{{\bf k}^2}\frac{d{\bf q}^2}{\pi{\bf q}^2}\int\limits_{z}^{1}dz^\prime\frac{\bar\alpha_S}{z^{\prime}}\Bigg)\; .
\end{equation}

The function ${\mathcal P}(z,{\bf k},{\bf q})$ is the gluon splitting function and in the original formulation and most of the literature~\mcite{CCFM1,*CCFM2,*CCFM3,*CCFM4,CASCADE1,*CASCADE2,Avsar:2009pf,*Avsar:2010ia} it takes this form
\begin{equation}\label{eq:orgsplf}
{\mathcal P}(z,{\bf k},{\bf q})=\frac{\Delta_{NS}(z,{{\bf k}}^2,|{\bf q}^{\prime}|)}{z}+\frac{1}{1-z}\;.
\end{equation}

Note that the finite terms, those without of a pole in $z=0$ or $z=1$, from the full gluon splitting function are neglected.
To study the small-$x$ asymptotic usually the term $1/(1-z)$ is neglected~\mcite{Avsar:2009pf,*Avsar:2010ia}
\begin{equation}\label{eq:1ozsplf}
{\mathcal P}(z,{\bf k},{\bf q})=\frac{\Delta_{NS}(z,{{\bf k}}^2,|{\bf q}^{\prime}|)}{z}\;.
\end{equation}

It turns out, that together with including the $1/(1-z)$ term, it is important and interesting to include the finite terms. We will include the $-2$ term and split it between the $1/z$ and $1/(1-z)$ term equaly as it was suggested in~\mcite{Andersson:2002cf,Jung:2003wu,*Hansson:2003xz}
\begin{equation}\label{eq:modsplf}
{\mathcal P}(z,{\bf k},{\bf q})=\Delta_{NS}(z,{{\bf k}}^2,|{\bf q}^{\prime}|)\frac{1-z}{z}+\frac{z}{1-z}\;.
\end{equation}

\subsection{The KGBJS equation}

We will follow the line of authors of~\mcite{Kutak:2012yr,*Kutak:2011fu} who suggested a non-linear extension of the CCFM equation~\eqref{eq:IS-CCFM} in this form
\begin{equation}\label{eq:IS-KGBJS}
\begin{split}
\mathcal{\tilde F}(x,{\bf k},p)&=\mathcal{\tilde F}_0({\bf k})\\&+{\bar\alpha}_S
\int\frac{d^2\bar{\bf q}^\prime}{\bar{\bf q}^{\prime 2}}\int\limits_{x}^{1-\frac{Q_0}{|{\bar{\bf q}}^{\prime}|}}\frac{dz}{z}\Big(\mathcal{\tilde F}(x/z,{\bf k}^{\prime},|\bar{\bf q}^{\prime}|)-\delta({{\bar{\bf q}}^{\prime 2}}-{\bf k}^2)({\bar{\bf q}}^{\prime 2})\mathcal{\tilde F}^2(x/z,\bar{\bf q}^{\prime},|\bar{\bf q}^{\prime}|)\Big)\\&\times\;\theta(p-z|{\bar {\bf q}}^\prime|){\mathcal P}(z,{\bf k},{\bf q})\Delta_S(p,z|\bar{\bf q}^{\prime}|,Q_0)\;.
\end{split}
\end{equation}

An important comment is required about the upper limit in the $z$ and $z^{\prime}$ integrals in~\eqref{eq:IS-CCFM} and~\eqref{eq:sud}. This upper limit regulates integrals with a pole in $z,z^{\prime}=1$. It is easy to see, that it also generates a limit on the value of the variable $|{\bf q}^\prime|>Q_0/(1-z)$ and also on the variable $|{\bar {\bf q}}|>Q_0/(1-z)$. The latter limit is going to affect transversal momentum dependence of the solution of the KGBJS equation~\eqref{eq:IS-KGBJS} near the soft scale $|{\bf k}|\sim Q_0$. In case when $Q_0<|{\bf k}|<Q_0/(1-z)$ the non-linear term in the~\eqref{eq:IS-KGBJS} will be equal to $0$ rendering the solutions of the linear and the non-linear equations equal. For $|{\bf k}|=Q_0$ we thus have
\begin{equation}\label{eq:ccfmmatchcond}
\mathcal{\tilde F}(x,Q_0,p)=\mathcal{F}(x,Q_0,p)\; .
\end{equation}

This is not what we would intuitively expect from a gluon Green function with growth tamed by a non-linear correction. We are going to study properties of the solutions of the non-linear equation~\eqref{eq:IS-KGBJS}, but we suggest a modification which can give a more natural behaviour of its solution. We modify the argument of the delta function in the non-linear term
\begin{equation}\label{eq:IS-KGBJSD}
\begin{split}
\mathcal{\tilde F}_D(x,{\bf k},p)=\mathcal{\tilde F}_0({\bf k})+&{\bar\alpha}_S
\int\frac{d^2\bar{\bf q}^\prime}{\bar{\bf q}^{\prime 2}}\int\limits_{x}^{1-\frac{Q_0}{|{\bar{\bf q}}^{\prime}|}}\frac{dz}{z}\Big(\mathcal{\tilde F}_D(x/z,{\bf k}^{\prime},|\bar{\bf q}^{\prime}|)\\-&\delta\big({{\bar{\bf q}}^{\prime 2}}-{\bf k}^2/(1-z)^2\big)({\bar{\bf q}}^{\prime 2})\mathcal{\tilde F}_D^2(x/z,\bar{\bf q}^{\prime},|\bar{\bf q}^{\prime}|)\Big)\\\times&\;\theta(p-z|{\bar {\bf q}}^\prime|){\mathcal P}(z,{\bf k},{\bf q})\Delta_S(p,z|\bar{\bf q}^{\prime}|,Q_0)
\end{split}
\end{equation}

to shift its 'pole' outside of the interval $\big(Q_0,Q_0/(1-z)\big)$.

\section{The CCFM -- BFKL correspondence}

A thorough analysis of the relation between the CCFM and the BFKL equations was studied in~\mcite{Avsar:2009pf,*Avsar:2010ia}. Here we are going to point out a very simple way from the CCFM to the BFKL equation.

Let us calculate the expressions for equations~\eqref{eq:sud} and~\eqref{eq:nsud} in a case of constant ${\bar\alpha}_S$. After performing the integrals one arrives to these formulae (for $z|{\bar{\bf q}}|>Q_0/(1-z)$):
\begin{equation}
\Delta_S(p,z|\bar{\bf q}|,Q_0)=\exp\Big(-{\bar\alpha}_S\ln\frac{p}{z|{\bar{\bf q}}|}\ln\frac{z|{\bar{\bf q}}|\,p}{Q_0^2}\Big)
\end{equation}

\begin{center} and \end{center}
\begin{equation}
\Delta_{NS}(z,{\bf k}^2,|{\bf q}|)=\exp\Big(-{\bar\alpha}_S\ln\frac{1}{z}\ln\frac{{\bf k}^2}{z{\bf q}^2}\Big)\; .
\end{equation}

It would be useful now to see how these form factors are related to the Regge form factor which is the result of ressumation in the BFKL equation. The answer on the question gives a lot of insight into the different properties and ways of working of the CCFM and the BFKL equations.

The relation between the CCFM and the BFKL equations is not a simple small-$x$ limit, but, understandably, includes it. The CCFM equation includes an extra scale $p$ which sets the hard scale of the process -- the maximal emission angle in the evolution. There is no such scale in the BFKL case -- every emission sets a semihard scale. Intuition tells us to perform limits $x\rightarrow 0\implies z\rightarrow 0\implies {\bar{\bf q}}\rightarrow {\bf q}\,$ and set $p\rightarrow |{\bf q}|$. As a result we get 
\begin{equation}
\Delta_S(p,(z|\bar{\bf q}|,Q_0)\Delta_{NS}(z,{\bf k}^2,|{\bf q}|)\xrightarrow[p\rightarrow |{\bf q}|]{z\rightarrow 0}\Delta_R(z,{\bf k}^2,Q_0)\;.
\end{equation}

In addition $\theta(p-z{\bf {\bar q}})\xrightarrow[p\rightarrow |{\bf q}|]{z\rightarrow 0}1$. By removing the $1/(1-z)$ term and dropping the ${\bf q}$ dependence in the Green function, we obtain exactly the BFKL equation (the linear part of \eqref{eq:BKresummed}) with $\mu=Q_0$. 

One of advantages of the knowledge of the relation between the equations is, for example, a possibility to cross check their solutions against each other.

\section{Numerical solutions of the equations: Discussion of the results}\label{sec:results}

We solve the equations~\eqref{eq:IS-CCFM},~\eqref{eq:BKresummed}, the BFKL equation and equations~\eqref{eq:IS-KGBJS} and~\eqref{eq:IS-KGBJSD} by iteration on a lattice under certain additional conditions:
\begin{itemize}
\item We set the ${\bar \alpha}_S=0.2$. The running $\alpha_S$ case will be discussed in future studies.

\item We will study the CCFM equation with and without the $1/(1-z)$ to estimate its importance and importance of the finite terms. We will choose different forms of the gluon splitting function listed in the section~\ref{sec:equations} (equations~\eqref{eq:orgsplf},~\eqref{eq:1ozsplf} and~\eqref{eq:modsplf}).

\item We have written the CCFM equation with a form of the Non-Sudakov form factor~\eqref{eq:nsud} which requires a kinematical constraint to satisfy unitarity. We will thus require $|{\bf q}|<|{\bf k}|/\sqrt{z}$ ~\mcite{Avsar:2009pf,*Avsar:2010ia} in the CCFM kernel. 

\item We will also impose the kinematical constraint on the BFKL and BK equations and compare the results with unconstrained ones.

\item We set the parameters $\mu=Q_0=1\;GeV$.

\item To mimic energy-momentum conservation we apply an upper limit on the ${\bar {\bf q}},\;{\bf q}<\sqrt{s_{tot}}$ integration. In this publication we choose $\sqrt{s_{tot}}\simeq(1\,GeV/x_{min})$. Where $x_{min}$ is the minimal momentum fraction $x$ for which we parametrise the solution of a given equation.

\item We take the initial condition to be
\begin{equation}
\tilde{f}^0({\bf k}^2)=\frac{C_{in}}{|{\bf k}|}
\end{equation}

with $C_{in}$ being a constant parameter. We set $C_{in}=0.1$ for the BFKL and the BK equations and $C_{in}=0.5$ for the CCFM and the KGBJS equations (in $\mathcal{F}_0({\bf k}^2)$ and $\mathcal{\tilde{F}}_0({\bf k}^2)$).
\end{itemize}

In next we are going to discuss the results of the numerical calculations.

\subsection{The BFKL, BK and BFKL with kinematical constraint}\label{sec:BFKLnum}

\begin{figure}[tbh]
\vspace{0.75cm}
  \begin{picture}(30,0)
    \put(22, -124){
      \includegraphics{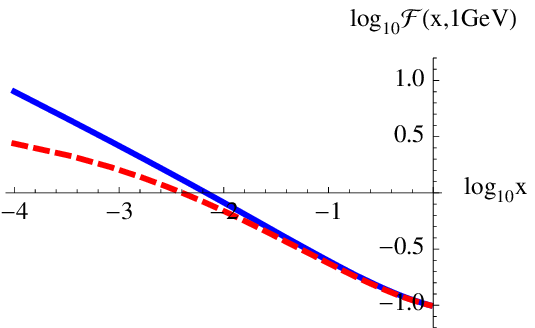}
    }
    \put(213, -138){
      \includegraphics{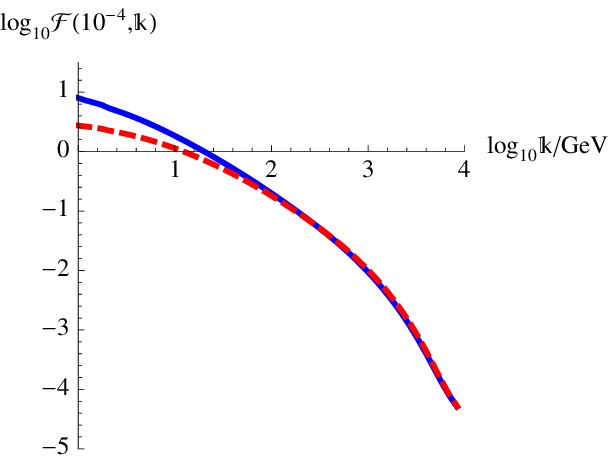}
    }

     \end{picture}
\vspace{4.7cm}
\caption{The $x$ and ${\bf k}$ distributions for BK -- red dashed line and BFKL -- blue solid line.}
\label{fig:plots1}
\end{figure}

In the figure~\ref{fig:plots1} we plot a solution of the BFKL equation compared with a solution of the BK equation for the same initial condition (in the beginning of the section~\ref{sec:results}). In the BK case we can see slowing down of the $1/x^{\lambda}$ growth of the BFKL Green function (on the left in the~\ref{fig:plots1}). 

Note that the transversal momentum distribution (on the right in the~\ref{fig:plots1}) does not follow a simple power behaviour, but falls fast near the limit for the ${\bf q}$ integration $\sqrt{s_{tot}}=10^4\;GeV$. The growth also slows down near the soft cut-off $Q_0=1\;GeV$.

\begin{figure}[tbh]
  \begin{picture}(30,0)
    \put(12, -154){
      \includegraphics{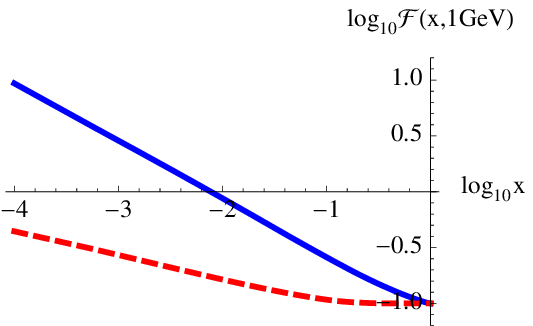}
    }
    \put(223, -168){
      \includegraphics{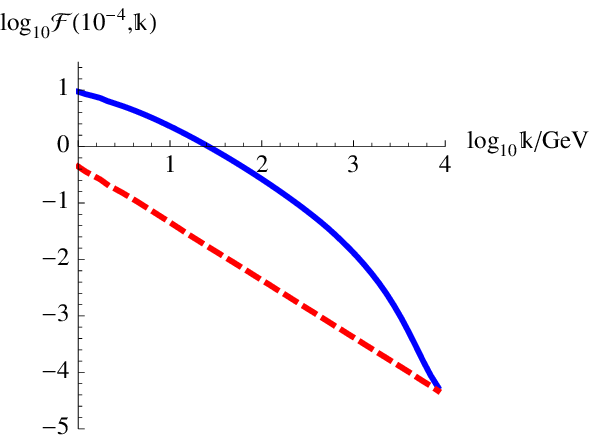}
    }

     \end{picture}
\vspace{5.7cm}
\caption{The $x$ and ${\bf k}$ distributions for BFKL -- blue solid line and BFKL with the kinematical constraint -- red dashed line.}
\label{fig:plots2}
\end{figure}

In the figure~\ref{fig:plots2} we compare the solutions of the BFKL equation without the kinematical constraint and with the kinematical constraint $|{\bf q}|<|{\bf k}|/\sqrt{z}$. We can see that the slope of the $x$ distribution is now less steep and also the distribution starts to grow only at around $x\sim10^{-1}$. The latter is a consequence of reduction of the integration region over ${\bf q}$ for large $z$.

The ${\bf k}$ spectrum of the solution of the BFKL equation with the kinematical constraint follows approximately a power distribution $\sim 1/|{\bf k}|$. For $|{\bf k}|=\sqrt{s_{tot}}$ the integration region over {\bf q} is the same with and without the kinematical constraint. As a consequence the ${\bf k}$ distributions have to match for $|{\bf k}|=\sqrt{s_{tot}}$ as seen in the plot in figure~\ref{fig:plots2}. Latter explains the sudden drop of the transversal momentum spectrum in the pure BFKL case, which, as we can now see, is a finite momentum effect.

The comparison of the kinematically constrained BFKL $x$ and ${\bf k}$ spectra with similar distributions obtained by applying kinematical constraint on the BK equation in figure~\ref{fig:plots2}, show the same pattern as for unconstrained BFKL and BK solutions. The suppression effect by the non-linear term is smaller, because the magnitude of the Green function is also smaller. In addition we can see, that the finite momentum effects are suppressed.

In next we are going to compare our results only with the solutions of the kinematically constrained BFKL equation, since the BFKL without the kinematical constraint is not complete and since the constraint is applied also in the CCFM equation.

\subsection{Different versions of the CCFM equation}

\begin{figure}[tbh]
\vspace{1.2cm}
  \begin{picture}(30,0)
    \put(10, -128){
      \includegraphics{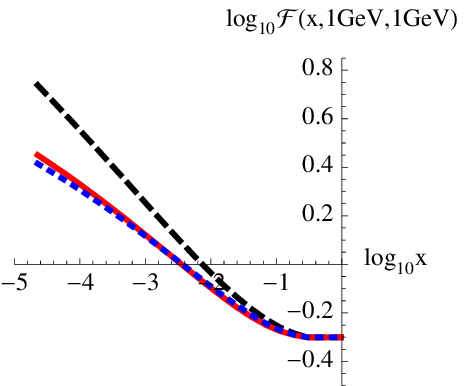}
    }
    \put(220, -120){
      \includegraphics{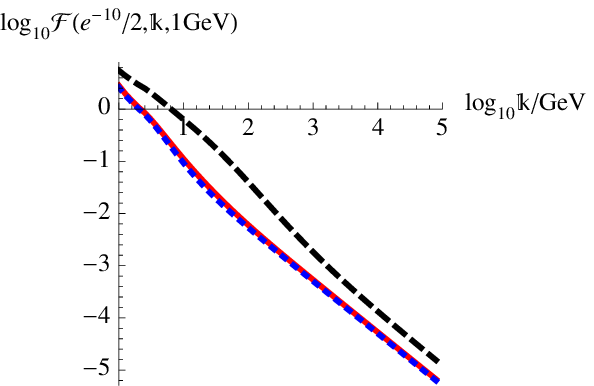}
      }
        \put(20, -285){
      \includegraphics{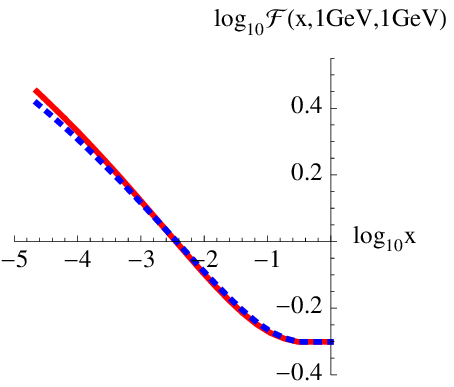}
    }    
        \put(220, -285){
      \includegraphics{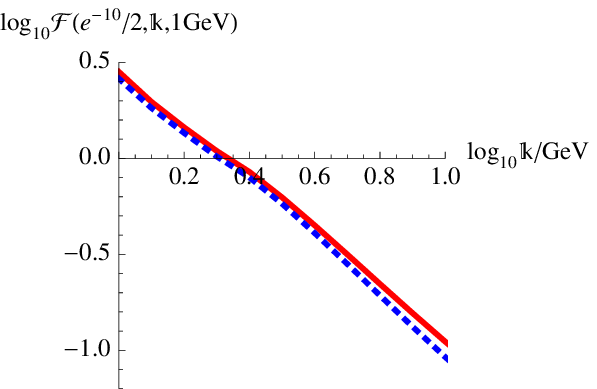}
    }        
     
     \end{picture}
\vspace{10.2cm}
\caption{The $x$ and ${\bf k}$ distributions of different versions of the CCFM equation. The lines correspond to the CCFM equation with the splitting function:~\eqref{eq:orgsplf} -- black dashed line,~\eqref{eq:1ozsplf} -- blue dotted line and~\eqref{eq:modsplf} -- solid red line.}
\label{fig:plots4}
\end{figure}

In plots in the figure~\ref{fig:plots4} we compare solutions of the CCFM equations which differ by choice of the splitting function. 

It is important to study especially the behaviour of the solution of the CCFM with the splitting function with only terms singular in $z=0$ and $z=1$~\eqref{eq:orgsplf}. The most interesting is the $x$ distribution. If we compare it with the $x$ distribution of the solution of the CCFM equation with splitting function~\eqref{eq:1ozsplf}, it seems that the inclusion of the $1/(1-z)$ term causes an enhancement in the small $x$ region.

Let us do a careful analysis of what happens at $z\rightarrow 0$ taking the splitting function~\eqref{eq:orgsplf}. Let us calculate the Sudakov form factor with fixed $p=|{\bf k}|=Q_0=1\;GeV$ to see, that for fixed $|{\bar {\bf q}}|$ it grows with $z$ getting smaller:
\begin{equation}
\Delta_S(1\;GeV,z|\bar{\bf q}|,1\;GeV)=\exp\Big(-{\bar\alpha}_S\ln\frac{1\;GeV}{z|{\bar{\bf q}}|}\ln\frac{z|{\bar{\bf q}}|}{1\;GeV}\Big)=\exp\Big({\bar\alpha}_S\ln^2\frac{1\;GeV}{z|{\bar{\bf q}}|}\Big)\;.
\end{equation}

In the case of the $1/z$ term is the enhancement compensated by the Non-Sudakov form factor. In the case of the second term in splitting function~\eqref{eq:orgsplf} $1/(1-z)\rightarrow 1$, but there is no suppression by a Non-Sudakov form factor. Just by writing down an estimation of the contribution of the $1/(1-z)$ term by the ${\bar{\bf q}}$ integral
\begin{equation}
\int\limits_{1}^{1/\sqrt{z}}\frac{d{\bar{\bf q}}}{{\bar{\bf q}}}\exp\Big({\bar\alpha}_S\ln^2\!\frac{1}{z|{\bar{\bf q}}|}\Big)
\end{equation}

we can clearly see that the integral over $|{\bar{\bf q}}|$ is going to grow fast for small $z$ and thus enhance the growth with $x\rightarrow 0$.

From the above estimations we can conclude, that the finite terms in the gluon splitting function, which were considered to give subleading contributions, have to play a key role in cancelling the anomalous growth. Indeed, looking at the plots in figure~\ref{fig:plots4}, we can see that there is no significant small $x$ enhancement, if we consider the splitting function~\eqref{eq:modsplf}. The term $1/(1-z)$ turns into $z/(1-z)$ vanishes in the limit $z\rightarrow 0$ effectively suppressing the growth. 

\begin{figure}[tbh]
  \begin{picture}(30,0)
    \put(12, -154){
      \includegraphics{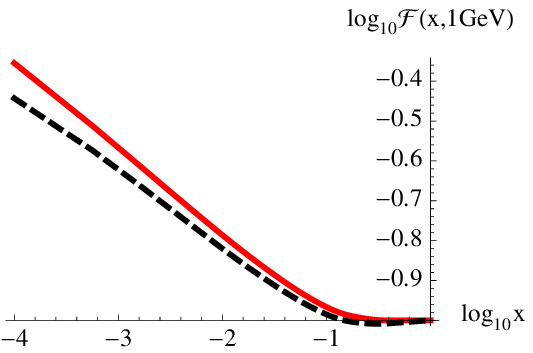}
    }
    \put(223, -168){
      \includegraphics{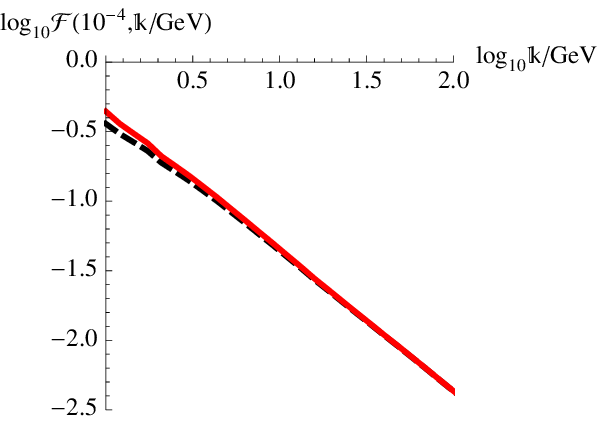}
    }

     \end{picture}
\vspace{5.7cm}
\caption{The $x$ and ${\bf k}$ distributions for BK -- black dashed line and BFKL -- red solid line with the kinematical constraint.}
\label{fig:plots3}
\end{figure}

However, what we observe is a small counter intuitive enhancement for small $x$ even after including the finite term. For large $x$ we see a suppression similar to the one observed for the BFKL with the kinematical constraint (figures~\ref{fig:plots2} and~\ref{fig:plots3}).

The ${\bf k}$ distributions of CCFM solutions with~\eqref{eq:1ozsplf} and~\eqref{eq:modsplf} splitting functions have a similar slope. The slope is somehow steeper than the one of the kinematically constraint BFKL equation (figure~\ref{fig:plots2} and~\ref{fig:plots3}) since it behaves like a power $\sim 1/|{\bf k}|^{1.5}$ till it reaches the magnitude of the initial condition at around $|{\bf k}|\sim 100\;GeV$. 

The difference in the steepness of the slopes of the BFKL and the CCFM equations could be the origin of differences between various observables obtained from their Green functions, such as the diffusion pattern and angular correlations, studied in~\mcite{Chachamis:2011rw}.

\subsubsection{The $p$ dependence}

\begin{figure}[tbh]
\vspace{1cm}
  \begin{picture}(30,0)
    \put(10, -128){
      \includegraphics{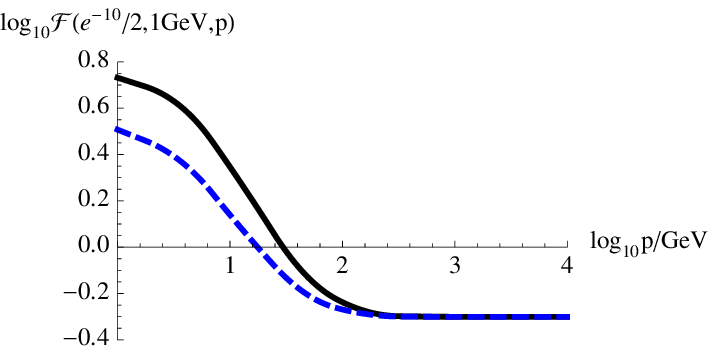}
    }
    \put(225, -130){
      \includegraphics{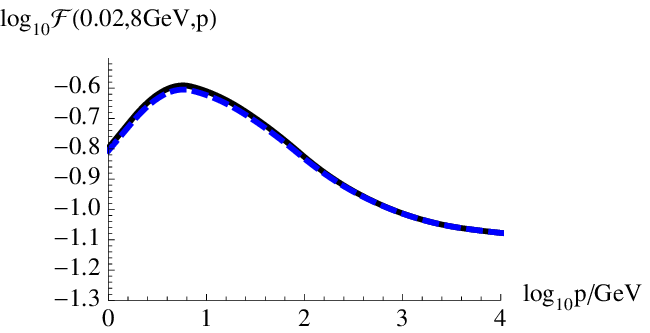}
      }
        \put(113, -305){
      \includegraphics{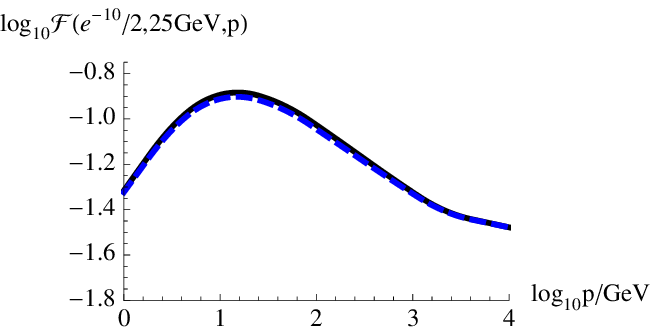}
    }

     \end{picture}
\vspace{11.2cm}
\caption{The $p$ dependence of the solution of the CCFM equation -- the black solid line -- compared with the modified KGBJS equation~\eqref{eq:IS-KGBJSD} -- the blue dashed line. For fixed $x$ and ${\bf k}$ with the splitting function~\eqref{eq:modsplf}.}
\label{fig:plots7}
\end{figure}

In the figure~\ref{fig:plots7} we plot the $p$ dependence of the the solution of the CCFM equation with the splitting function~\eqref{eq:modsplf} for 3 different fixed values 
of $x$ and ${\bf k}$.

Each of the distributions in figure~\ref{fig:plots7} peaks at $p\simeq|{\bf k}|$. This is because when $|{\bf k}|\simeq p$ the biggest contribution to the integral on the right hand side of the CCFM equation comes from the region in which ${\bar {\bf q}}$ is small. One can explain the distributions using a simple physical picture. If $p$ represents the hard scale in a form of transversal momentum, then in the case when $p\simeq |{\bf k}|$ the emitted transversal momentum $|{\bf q}|\simeq |{\bar{\bf q}}|$ is small and there is an enhancement by a factor $1/|{\bar{\bf q}}|$.

There is also phase space enhancement for the ${\bar{\bf q}}$ integration due to $\theta(p-z|{\bar{\bf q}}|)$ for $p< |{\bf k}|$ in the CCFM kernel~\eqref{eq:IS-CCFM}.
Due to the kinematical constraint there is no phase space enhancement for $p>|{\bf k}|$, but only suppression by the Sudakov form factor which explains the fall of the Green function.

\subsection{The KGBJS equation and its modifications}
 
\begin{figure}[tbh]
\vspace{1.2cm}
  \begin{picture}(30,0)
    \put(10, -128){
      \includegraphics{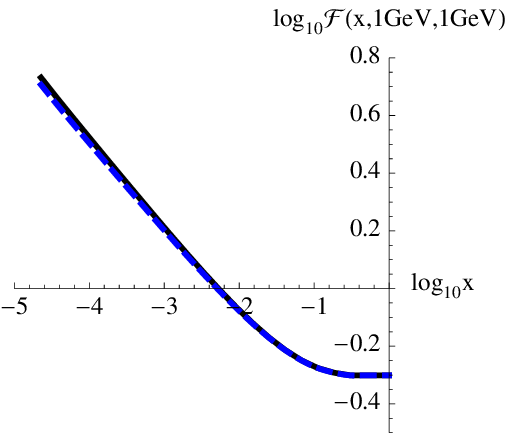}
    }
    \put(220, -120){
      \includegraphics{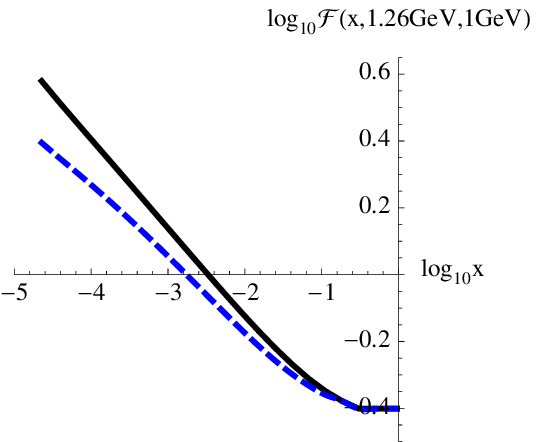}
      }
        \put(-10, -285){
      \includegraphics{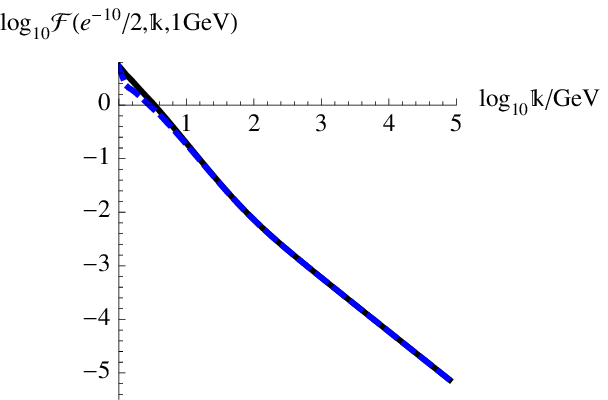}
    }    
        \put(210, -285){
      \includegraphics{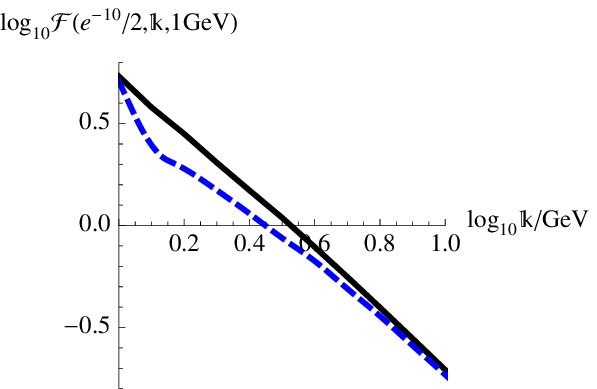}
    }        
     
     \end{picture}
\vspace{10.2cm}
\caption{The $x$ and ${\bf k}$ distributions of different versions of the CCFM equation -- the solid black line compared with the KGBJS equation~\eqref{eq:IS-KGBJS} -- the blue dashed line. Both of them with the splitting function~\eqref{eq:modsplf}.}
\label{fig:plots5}
\end{figure}

\begin{figure}[tbh]
\vspace{1.2cm}
  \begin{picture}(30,0)
    \put(10, -128){
      \includegraphics{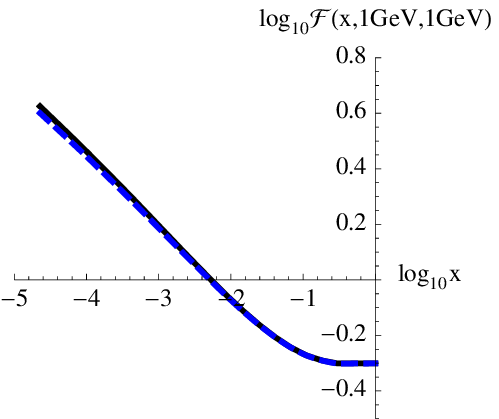}
    }
    \put(220, -120){
      \includegraphics{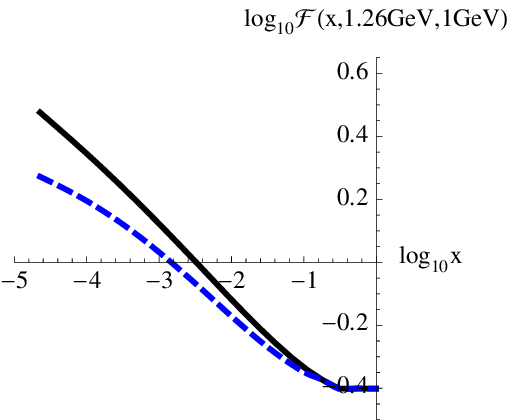}
      }
        \put(-10, -285){
      \includegraphics{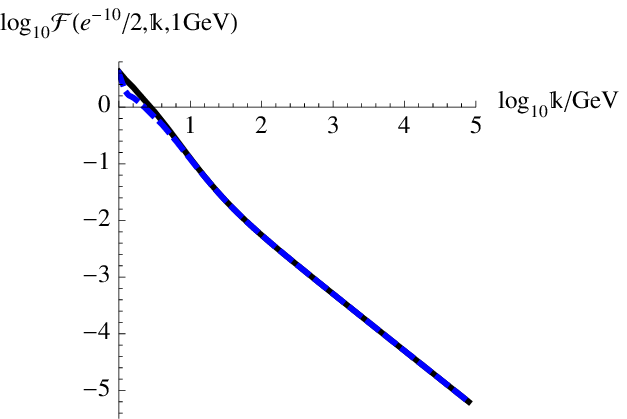}
    }    
        \put(210, -285){
      \includegraphics{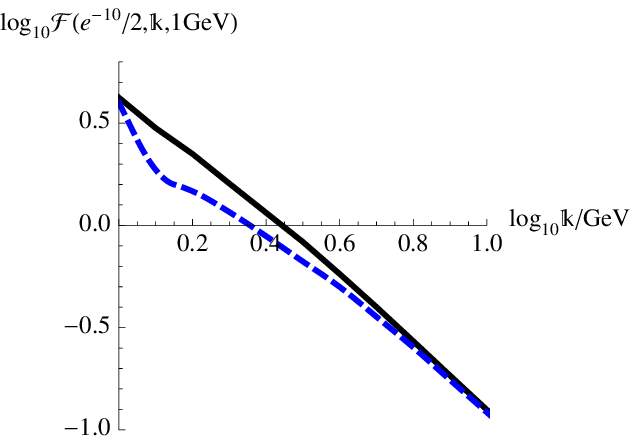}
    }        
     
     \end{picture}
\vspace{10.2cm}
\caption{The $x$ and ${\bf k}$ distributions of different versions of the CCFM equation -- the solid black line compared with the KGBJS equation~\eqref{eq:IS-KGBJS} -- the blue dashed line. Both of them with the splitting function~\eqref{eq:1ozsplf}.}
\label{fig:plots6}
\end{figure}

Let us look now at figures~\ref{fig:plots5} and~\ref{fig:plots6}. It is better to look first on the ${\bf k}$ distributions. We observe the predicted matching of the CCFM and the KGBJS solution at $|{\bf k}|=Q_0=1\;GeV$. For $|{\bf k}|>Q_0=1\;GeV$ there is a suppression in a form of a dip. The dip 'lasts' up to value $|{\bf k}|\simeq 10\;GeV$.

In the point $|{\bf k}|=Q_0=1\;GeV$ there is almost no difference between the $x$ distributions of the CCFM and the KGBJS solutions. Different situation is in $|{\bf k}|=Q_0=1.26\;GeV$ where we observe similar suppression of growth as for the BK equation, figures~\ref{fig:plots1} and~\ref{fig:plots3}.

All the observed effects are similar for both choices of the splitting function~\eqref{eq:1ozsplf} and~\eqref{eq:modsplf} (figures ~\ref{fig:plots5} and~\ref{fig:plots6}).

\subsection{The modified KGBJS equation}

In this subsection we are going to discuss the comparison of the solution of the CCFM equation~\eqref{eq:IS-CCFM} with our new modified KGBJS equation~\eqref{eq:IS-KGBJSD} both with the splitting function~\eqref{eq:modsplf}.

\begin{figure}[tbh]
\vspace{1cm}
  \begin{picture}(30,0)
    \put(10, -128){
      \includegraphics{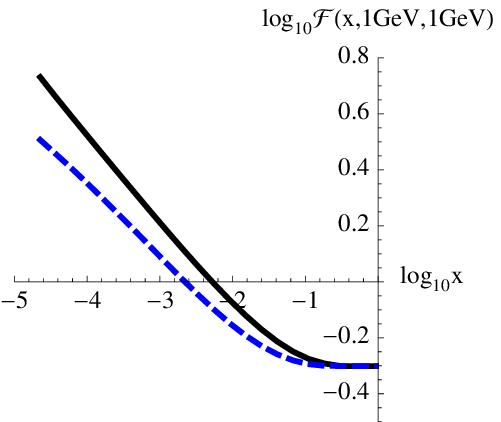}
    }
    \put(225, -120){
      \includegraphics{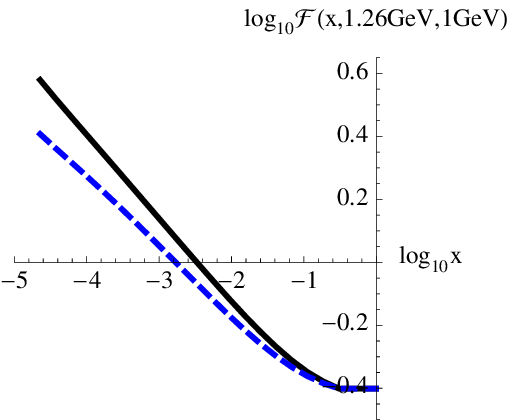}
      }
        \put(-5, -315){
      \includegraphics{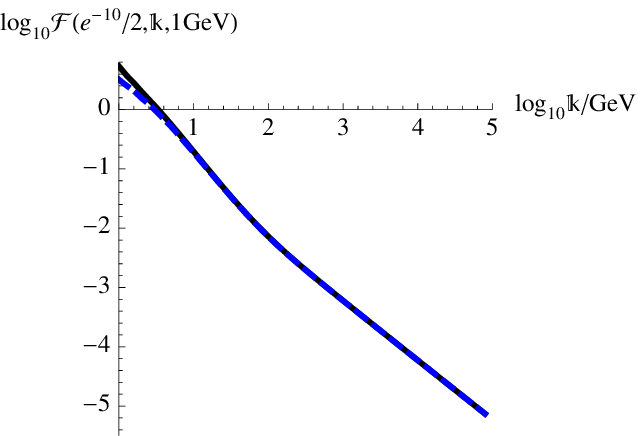}
    }    
        \put(215, -315){
      \includegraphics{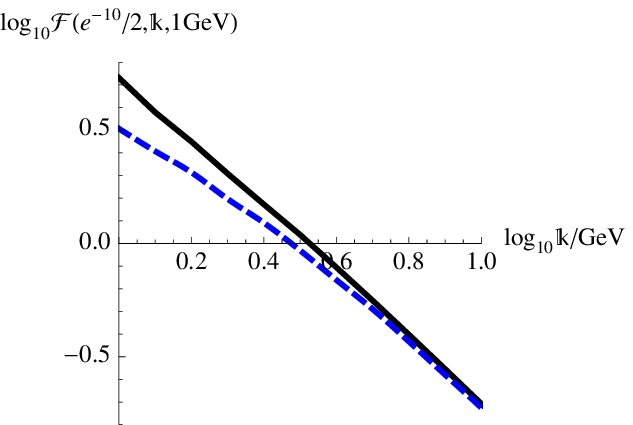}
    }    
     
     \end{picture}
\vspace{11.2cm}
\caption{The $x$ and ${\bf k}$ distributions of different versions of the CCFM equation -- the solid black line -- compared with the modified KGBJS equation~\eqref{eq:IS-KGBJSD} -- the blue dashed line. Both of them with the splitting function~\eqref{eq:modsplf}.}
\label{fig:plots8}
\end{figure}

Let us take a look on the plots in the figure~\ref{fig:plots8}. We will first discuss the ${\bf k}$ distributions. We can see that the modification really breaks the condition~\eqref{eq:ccfmmatchcond}, so the solutions of the linear equation and the non-linear equation are not equal at the soft cut-off $Q_0$. We do not observe a formation of the dip as for the original KGBJS equation (figures~\ref{fig:plots5} and~\ref{fig:plots6}). The behaviour of the suppression for small ${\bf k}$ is similar to the one generated by the BK equation (figures~\ref{fig:plots1} and~\ref{fig:plots3}).

The natural behaviour of the solution of~\eqref{eq:IS-KGBJSD} is reflected also in the $x$ distributions. The difference between $x$ distribution of the CCFM and the non-linear equation gets smaller for $|{\bf k}|$ getting bigger.

We have also compared the $p$ dependence of above mentioned solutions plotted in the figure~\ref{fig:plots7}. Except of suppression for the non-linear equation, which is getting smaller with bigger $|{\bf k}|$ and $x$, there is no significant difference in the $p$ dependence of the solutions.

\section{Summary and Conclusions}

We have numerically solved the BFKL and the BK equations with and without the kinematical constraint. We have obtained solutions of different versions of the CCFM equation. We have also solved the KGBJS equation~\mcite{Kutak:2012yr,*Kutak:2011fu} and its different modifications.

We have studied the transversal momentum ${\bf k}$ and $x$ distributions of the obtained solutions analytically and also numerically.

We confirm the previous suggestions to include finite terms into the gluon splitting function used in the CCFM equation and find their inclusion very important.

We find that solutions of the CCFM and the KGBJS equations match at the soft cut-off which implies no suppression in the point where the CCFM solution has the biggest magnitude.

We suggest a modification of the KGBJS equation which removes the non-physical behaviour of the solutions of the original equation near the soft cut-off. We demonstrate the improvement by a numerical solution of the modified equation.

The resulting suppression due to the non-linear term in the solution of the new equation~\eqref{eq:IS-KGBJSD} is a result of complicated interplay between values of the Green function in the small $x$ and also large $x$ phase space regions.

Although the investigation presented here shows, that it is not easy to find a natural model for a non-linear extension of the CCFM equation, we recommend the improved equation~\eqref{eq:IS-KGBJSD} to be a subject of more studies of inclusive and exclusive observables.

\section*{Acknowledgements}

The author of this publication would like to thank N\'estor Armesto and Carlos Salgado for giving him an opportunity to work in their group and Krzysztof Kutak, Guillaume Beuf and Hannes Jung for very useful and inspiring discussions.

\bibliographystyle{JHEP}
\bibliography{deak_michal}

\newpage

\end{document}